\documentclass[onecolumn,draftcls,12pt, twoside]{IEEEtran}
\pdfoutput=1
\usepackage[usenames, dvipsnames]{color}
\usepackage[cmex10]{amsmath} 
\usepackage{amssymb,amsthm}

\usepackage[nolist]{acronym}

\usepackage{enumerate}

\usepackage{subcaption}
\usepackage{calc}
\usepackage{cite}
\usepackage{graphicx}

\usepackage{epstopdf}

\usepackage{array,multirow}

\usepackage{cite}
\usepackage{tikz,pgfplots}
\usepackage{comment}

\usepackage{mathtools}

\usepackage[bookmarks=false]{hyperref}

\usepackage{amsmath} 

\newcommand{\tf}{d}
\newcommand{\cri}{l}

\newcommand{\nuser}{n}

\newcommand{\ecri}{L}

\begin{acronym}
\acro{AP}{access point}
\acro{RA}{random access}
\acro{LT}{Luby Transform}
\acro{BP}{belief propagation}
\acro{i.i.d.}{independent and identically distributed}
\acro{IoT}{Internet of things}
\acro{MTC}{machine type communication}
\acro{PER}{packet error rate}
\acro{SIC}{successive interference cancellation}
\acro{URLLC}{ultra-reliable low latency communication}
\acro{PMF}{probability mass function}
\acro{CRI}{collision resolution interval}
\acro{PGF}{probability generating function}
\acro{CPGF}{conditional probability generating function}
\acro{EGF}{exponential generating function}
\acro{TGF}{transformed generating function}
\acro{rhs}{right-hand side}
\acro{CSA}{coded slotted ALOHA}
\acro{MPR}{multi-packet reception}
\acro{MTA}{modified binary-tree algorithm}
\acro{SICTA}{binary tree-algorithm with SIC}
\acro{BTA}{binary tree-algorithm}
\acro{STA}{standard tree-algorithm}
\acro{MST}{maximum stable throughput}
\acro{MTA}{modified tree-algorithm}
\acro{CRP}{collision-resolution protocol}
\acro{CAP}{channel-access protocol}
\acro{IRSA}{Irregular Repetition Slotted ALOHA}
\acro{CSA}{Coded Slotted ALOHA}

\acro{r.v.}{random variable}
\end{acronym}

\title{On d-ary tree algorithms with successive interference cancellation}

\author{\IEEEauthorblockN{Yash Deshpande\IEEEauthorrefmark{1}, \v Cedomir Stefanovi\' c\IEEEauthorrefmark{2}, H. Murat G\"ursu\IEEEauthorrefmark{3}, Wolfgang Kellerer\IEEEauthorrefmark{1} }\\
\IEEEauthorblockA{
\IEEEauthorrefmark{1}Chair of Communication Networks, Technical University of Munich, Germany \\
\IEEEauthorrefmark{2}Department of Electronic Systems, Aalborg University, Denmark\\
\IEEEauthorrefmark{3}Nokia Bell Labs, Munich, Germany\\
Email: \{yash.deshpande,wolfgang.kellerer\}@tum.de },cs@es.aau.dk,murat.gursu@nokia-bell-labs.com}

\begin{document}

\maketitle


\thispagestyle{empty}
\pagestyle{empty}

\begin{abstract}
In this paper, we outline the approach for the derivation of the length of the collision resolution interval for $d$-ary tree algorithms (TA) with gated access and successive interference cancellation (SIC), conditioned on the number of the contending users.
This is the basic performance parameter for TA with gated access.
We identify the deficiencies of the analysis performed in the seminal paper on TA with \ac{SIC} by Yu and Giannakis, showing that their analysis is correct only for binary splitting, i.e. for $d=2$.
We also provide some insightful results on the stable throughput that can be achieved for different values of $d$. 
\end{abstract}






\section{Calculation of the conditional length of the collision resolution interval}
\label{sec:d-ary}

\begin{figure}
\centering
     \begin{subfigure}[b]{0.3\textwidth}
         \centering
         \includegraphics[width=\textwidth]{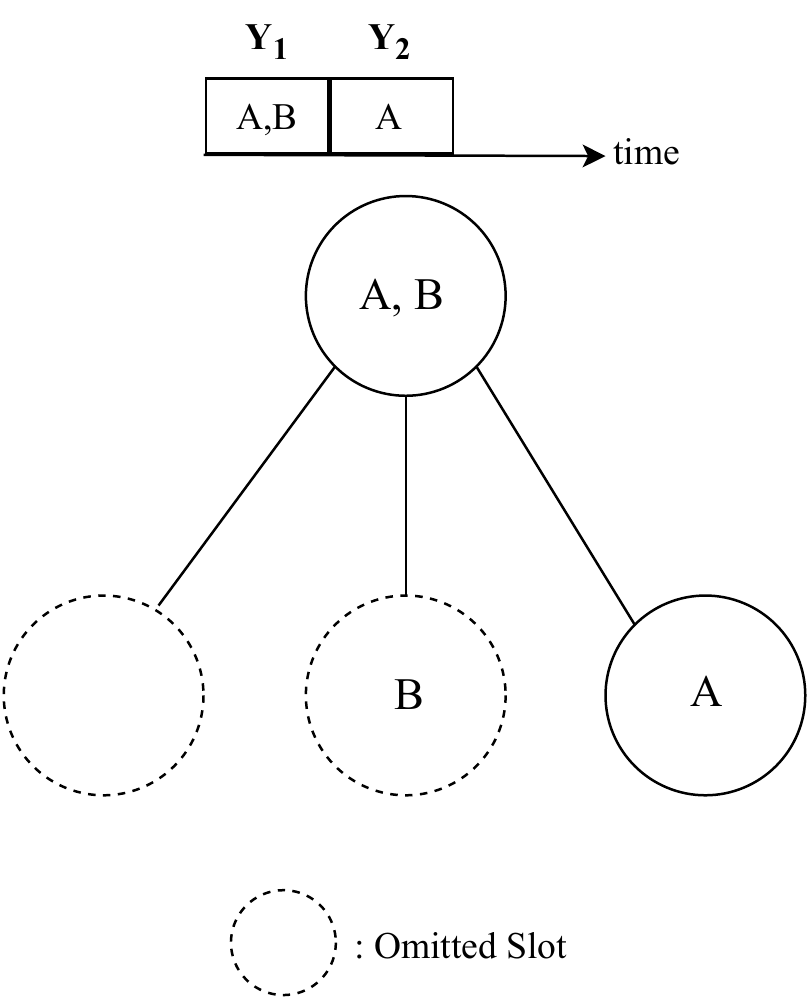}
         \caption{}
         \label{fig:sic_example}
     \end{subfigure}
     \hspace{4cm}
     \begin{subfigure}[b]{0.3\textwidth}
         \centering
         \includegraphics[width=\textwidth]{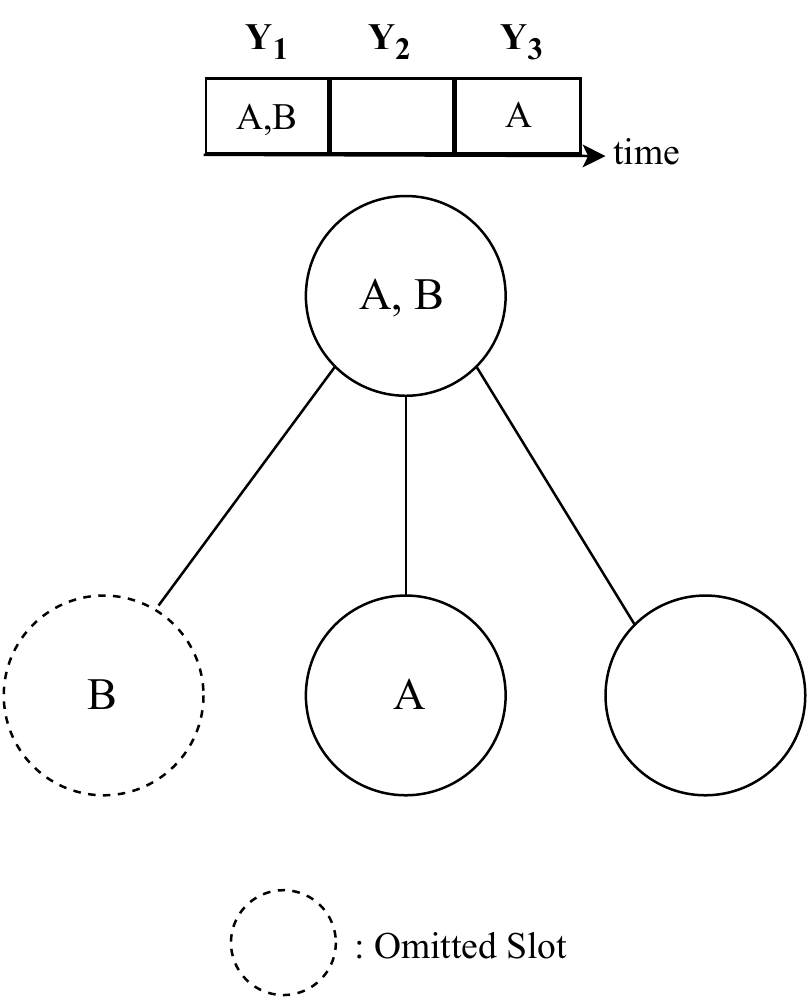}
         \caption{}
         \label{fig:sic_example_1}
     \end{subfigure}
  \caption{Example of a collision resolution with $n=2$ and $d=3$: (a) The CRI is 2 slots when there is single node in the 1st sub-group. According to \cite{SICTA}, this CRI should have been 3 slots. (b) The case when the CRI is indeed 3 slots. The authors in \cite{SICTA} did not consider scenario (a).}
  \label{fig:sic_example_both}
\end{figure}

Denote by $l_n$ the length (in slots) of the collision resolution interval (CRI) of a tree algorithm (TA) with gated access, conditioned on the initial number of contending users $n$.
This is the basic performance parameter of a TA, from which other performance parameters can be derived.
The approach for calculation of $l_n$ for $\tf$-ary tree algorithms with gated access and \ac{SIC} was presented in \cite[Section IV-A]{SICTA}. 
However, the premise of the analysis is incorrect when $d > 2$. 
The premise states the following (verbatim):
\begin{align}
\label{eq:d-ary_wrong}
    \cri_\nuser = \begin{cases} 1, & \text{if } n = 0,1 \\
    \sum_{j=1}^d \cri_{I_j} , & \text{if } n \geq 2
    \end{cases}
\end{align}
where $I_j$ is the number of nodes that select the $j$-th group, $j \in \{ 1, 2, \dots, d\}$.
We illustrate the shortcomings of the analysis through a simple example with $\nuser=2$, depicted in Fig.~\ref{fig:sic_example_both}. This example is similar to the example in \cite[Fig. 3]{SICTA}, only with $d=3$ instead of 2. In Fig.~\ref{fig:sic_example}, node~A selects the first group and node~B selects the second group after a collision in slot 1.
The receiver is able to decode the transmission occurring in slot 2 and after applying \ac{SIC}, is also able to recover the remaining transmission in slot 1.
As these two transmissions are the only ones in this tree, there is no need for further splitting, and the total duration of the \ac{CRI} is 2 slots.
However, according to \eqref{eq:d-ary_wrong}, the \ac{CRI} length in this example should be
\begin{align}
    l_2 = l_1 + l_1 + l_0 = 3.
\end{align}
Note that the CRI would also have been 2 slots if node B would have selected the third group.
In fact, if $\nuser=2$ and the first group has a single node, like in the Fig.~\ref{fig:sic_example}, the length of the CRI will be 2 slots, irrespective of the value of the splitting factor $\tf$ (given that $\tf \geq 2$).
On the other hand, the example in Fig.~\ref{fig:sic_example_1} shows the case when the \ac{CRI} length is indeed 3 slots and agrees with the formula \eqref{eq:d-ary_wrong}. This happens only when no collided node selects the first group.

In effect, the subsequent analysis in~\cite{SICTA} that exploited the results known for standard TA's with gated access (i.e., without SIC) becomes invalidated.
Specifically, the elegant conclusion drawn in \cite{SICTA} (verbatim)
\begin{align}
    (d-1) ( \ecri'_\nuser - 1) = d ( \ecri_\nuser - 1 )
\end{align}
where  $\ecri'_\nuser$ is the expected conditional length of the \ac{CRI} for the standard TA with gated access~\cite{capetanakis1979tree} and $\ecri_\nuser = \mathrm{E} [ l_n ]$ does not hold in general for $d > 2$.

Generalizing the insights shown above, we write
\begin{align}
\label{eq:dary_correct}
    \cri_\nuser = \begin{cases} 1, &  n = 0,1 \\
    \sum_{j=1}^{d_\text{min}} \cri_{I_j} , &  n \geq 2  
    \end{cases}
\end{align}
where $d_\text{min}$ is the minimum value of $o \in \{1,\dots, d\}$ for which the following holds
\begin{align}
\sum_{j=1}^{o} I_j \geq \nuser - 1.
\end{align}
The explanation of \eqref{eq:dary_correct} is intuitively clear -- the splitting process will stop as soon as there is a single user remaining from the original collision, no matter how many groups are left.

Unfortunately, when $d>2$, the expression \eqref{eq:dary_correct} can not be computed in the same manner as it can be done when $d = 2$.
In particular, the summands in \eqref{eq:dary_correct} are subject to the same recursion that holds for $\ecri_\nuser$, making the overall computation quickly intractable.
Further investigations with respect to this problem are out of the scope of this paper.

\section{Remarks on \ac{MST} of SICTA with gated access}

\begin{figure}[t]
  \centering
 \includegraphics[width=0.65\columnwidth]{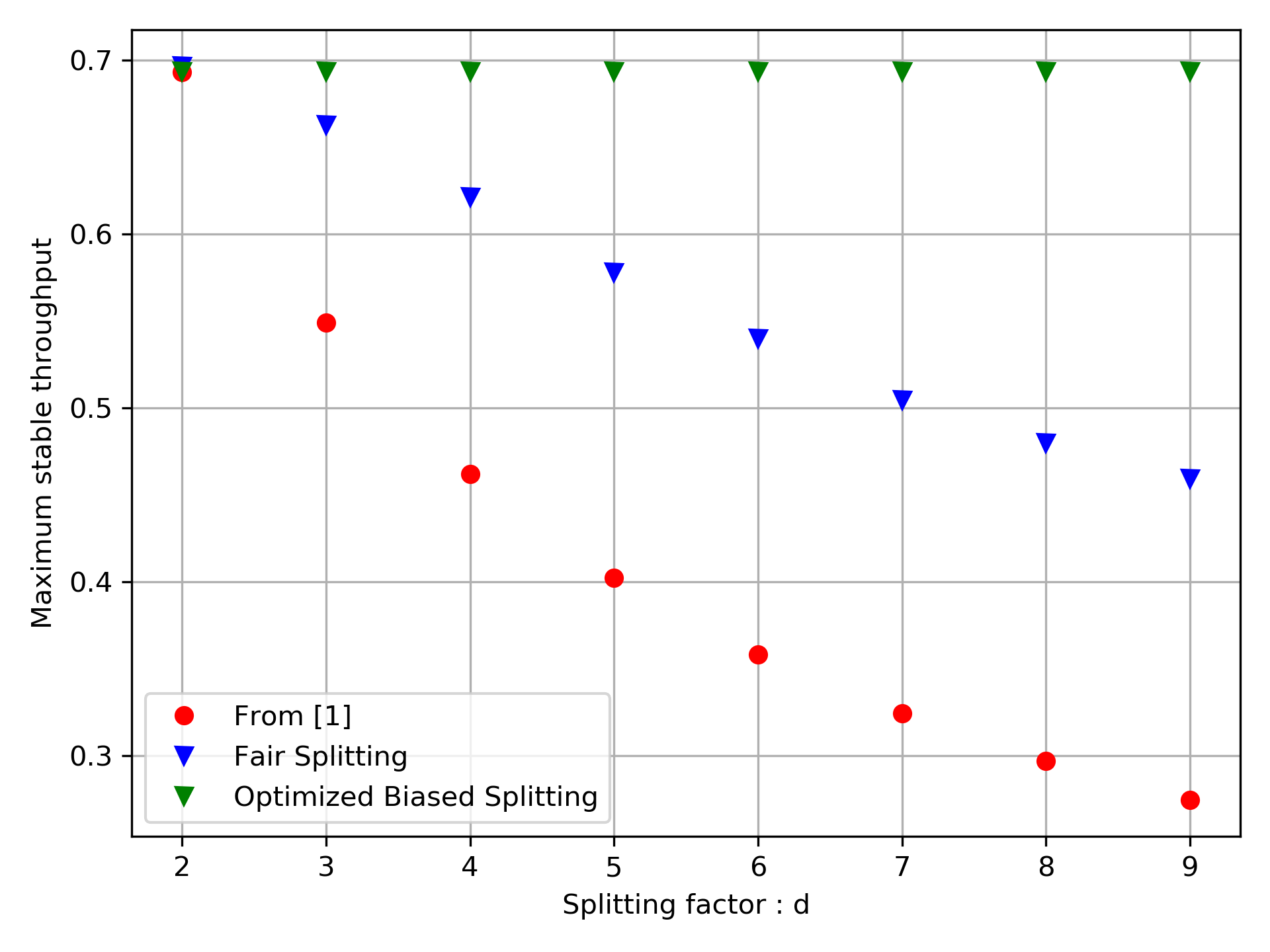}
  \caption{Throughput performance of $\tf$-ary tree-algorithms with \ac{SIC} as function of splitting factor $d$: red dots represent \ac{MST} results from~\cite{SICTA}, blue and green triangles represent the mean throughput obtained via simulations for $n=1000$ averaged over 10000 simulation runs with fair splitting and optimized biased-splitting, respectively.}
  \label{fig:sicta}
\end{figure}

The paper~\cite{SICTA} also provides results for \ac{MST} of $d$-ary SICTA with gated access, claiming (i) that fair splitting is the optimal choice for any $d \geq 2$, and (ii) that binary fair splitting achieves the highest \ac{MST}, see \cite[Section IV-B]{SICTA}.
However, these result are consequences of the wrong premise, as elaborated in Section~\ref{sec:d-ary}. 

We disprove these claims via Fig.~\ref{fig:sicta}, which compares the results for \ac{MST} with the gated access presented in~\cite{SICTA}, shown in \cite[Fig 6.]{SICTA}, with some illustrative throughput results obtained through simulations by measuring the actual CRI length $\ecri_n$ and computing $n / \ecri_n$.
In each simulation the number of nodes is $\nuser=1000$ and the results are obtained by averaging over 10000 simulation runs.
Such computed results for a high number of nodes $\nuser$ tend to serve as a proxy for the throughput for gated access,
\begin{equation}
    T = \frac{1}{\lim_{\nuser \rightarrow \infty} \ecri_n / n}
    \label{eq:proxy_mst}
\end{equation}
see \cite[Section IV-B]{SICTA}, where $T$ denotes the throughput.

Evidently, the results for $d>2$  are wrong, even for the fair splitting case.
In particular, in \cite{SICTA} the authors claim that $T \approx \ln(d) / d - 1$; these values are depicted by the red dots in Fig.~\ref{fig:sicta}.
However, there is a huge difference between the red curve and the blue one, which corresponds to the fair-splitting results obtained via simulations using the corrected premise.
The discrepancies among the curves become more pronounced as $\tf$ increases. 

The figure also shows that when an optimized biased-splitting is applied for $d > 2$, the numerically obtained throughput has the same value\footnote{Disregarding the precision loss due to the averaging over a finite number of realizations.} as the MST obtained for $d = 2$, which is $\ln 2$.
For the optimized splitting probabilities we used the following values
\begin{equation}
    p_{j} = \begin{cases}
    0.5^{j} & \; j \in \{1,..,d-1 \} \ \\
    0.5^{d-1} & \; j = d
    \end{cases}
\end{equation}
where $p_j$ is the probability that a node selects the j-th group,  $j \in \{1,..,d\}$.

In summary, these results shows that the biased splitting outperforms the fair one for $d > 2$, and that the claim that the highest MST is achieved exclusively for $d=2$ can not be assumed to hold anymore.









\bibliographystyle{IEEEtran}

\bibliography{main}

\end{document}